\documentclass[a4paper,11pt]{article}
\pdfoutput=1 

\usepackage{jheppub} 

\usepackage{mathrsfs}

\usepackage{graphicx}

\usepackage[usenames,dvipsnames]{color}
\usepackage{amsmath}
\usepackage{bbm}
\usepackage{amsfonts}
\usepackage{amssymb}
\usepackage{latexsym}
\usepackage{graphicx}
\usepackage[english]{babel}
\usepackage{multirow}
\usepackage{float}
\usepackage{url}
\usepackage{slashed}
\usepackage{xcolor}

\newcommand{\be}{\begin{equation}}
\newcommand{\ee}{\end{equation}}
\newcommand{\ba}{\begin{array}}
\newcommand{\ea}{\end{array}}
\newcommand{\bea}{\begin{eqnarray}}
\newcommand{\eea}{\end{eqnarray}}

\title{Enhanced Long-Lived Dark Photon Signals at the LHC}

\author[a]{Mingxuan Du,} 
\author[a,b,c,d]{Zuowei Liu}
\author[a]{and Van Que Tran}

\affiliation[a]{Department of Physics, Nanjing University, Nanjing 210093, China}
\affiliation[b]{Center for High Energy Physics, Peking University, Beijing 100871, China}
\affiliation[c]{Nanjing Proton Source Research and Design Center, Nanjing 210093, China} 
\affiliation[d]{CAS Center for Excellence in Particle Physics, Beijing 100049, China}

\emailAdd{mg1722004@smail.nju.edu.cn} 
\emailAdd{zuoweiliu@nju.edu.cn} 
\emailAdd{vqtran@nju.edu.cn}

\abstract{

We construct a model in which the standard model is extended by a hidden sector with two gauge $U(1)$ bosons. A Dirac fermion $\psi$ charged under both $U(1)$ fields is introduced in the hidden sector which can be a subcomponent of the dark matter in the Universe. Stueckelberg mass terms between the two new gauge $U(1)$ fields and the hypercharge gauge boson mediate the interactions between the standard model sector and the hidden sector. A remarkable collider signature of this model is the enhanced long-lived dark photon events at the LHC than the conventional dark photon models; the long-lived dark photons in the model can be discriminated from the background by measuring the time delay signal in the precision timing detectors which are proposed to be installed in the LHC upgrades and have an ${\cal O} (10)$ pico-second detection efficiency. Searches with current LHCb data are also investigated. Various experimental constraints on the model including collider constraints and cosmological constraints are also discussed.

}

\begin{document}
\maketitle
\flushbottom

\section{Introduction}

Recently, particles with a long lifetime have been 
studied extensively at colliders 
(see e.g.\ Ref. \cite{Alimena:2019zri} for a review). 
At the large hadron collider (LHC), 
the long-lived particles have been searched for 
in channels with displaced dilepton vertices 
\cite{Aad:2015rba, Aaboud:2018jbr, CMS:2014hka, Khachatryan:2014mea, Aad:2019tcc} 
and in channels with displaced jets vertices 
\cite{Aad:2012zn, Chatrchyan:2012sp, Aad:2013gva, Khachatryan:2015jha, Aaboud:2016dgf, Khachatryan:2016sfv, Aaboud:2017iio, Aaboud:2019opc, CMS:2016azs}. 
Searches for displaced vertices of collimated leptons or light hadrons with low $p_T$, 
which originate from light neutral particles such as dark photons, 
have been carried out at the LHC, 
e.g.\  by the ATLAS collaboration  
 \cite{Aad:2014yea, Aad:2019tua}.

Typically, for a particle to be considered as a long-lived particle at the LHC, 
the decay length has to be larger than ${\cal O}(1)$ mm 
so that the displaced vertex can be detected by the spatial resolution 
of the LHC detectors. 
Thus, the coupling strength between long-lived particles and 
the standard model (SM) particles is usually significantly reduced.  
For example, an electrophilic vector boson $A'$ that 
couples with electron via 
$g A'_\mu \bar e \gamma^\mu e$ has 
a decay width $\Gamma = g^2 m_{A'}/(12\pi)$ 
and a decay length $d = \gamma v \tau$, 
where 
$v$ ($\tau$) is the velocity (lifetime) of the vector boson $A'$,  
and $\gamma$ is the Lorentz boost factor. 
For such a particle to have a macroscopic 
decay length so that it can  give rise to a long-lived 
particle signature at the LHC, the decay coupling 
is typically small. 
For example, consider a typical decay length 
as $d \simeq 1$ m, one has  $g \sim {\cal O}(10^{-6})$ 
for the case where $m_{A'} = 1$ GeV and $\gamma = 100$. 
Thus simple long-lived vector boson models 
usually lead to a suppressed LHC cross section due to the 
small coupling constant needed for the large  
decay length.

In this paper, we construct a model that predicts a long-lived 
dark photon (LLDP) with a GeV scale mass.   
Unlike many other dark photon models, 
the production cross section of the GeV LLDP in our model 
at the LHC is not suppressed. 
This is because the production process of the LLDP is different 
from its decay process. 
We use the Stueckelberg mechanism 
to mediate the interaction between the hidden sector 
and the SM sector;  
the production and decay processes of the LLDPs are mediated 
by different Stueckelberg mass terms.

Recently, models in which LLDPs can have 
a sizable collider signal have been proposed   
in the literature. 
For example, Ref.\ \cite{Buschmann:2015awa} introduced a   
second boson with couplings to both SM quarks and the hidden fermion 
to produce dark photons at colliders. 
Ref.\ \cite{Arguelles:2016ney} used 
a dimension-five operator between a scalar $SU(2)_L$ triplet, 
the SM $SU(2)_L$ gauge bosons and the dark gauge boson 
to generate a non-abelian kinetic mixing term, which  
can enhance the LLDP signal. 
Potential large LLDP collider signals 
can also arise 
via top-partner decays \cite{Kim:2019oyh}, 
or via a Higgs portal interaction 
to the hidden QED \cite{Krovi:2019hdl}.

\section{The model}
\label{sec:model}

We introduce two Abelian gauge groups in the hidden sector: 
$U(1)_F$ with gauge boson $X^{\mu}$ and 
$U(1)_W$ with gauge boson $C^{\mu}$. 
We use the Stueckelberg mechanism 
to provide masses 
to the two new gauge bosons in the hidden sector, 
and also to mediate the interactions 
between the hidden sector and the 
SM sector  
\cite{Kors:2005uz, 
Feldman:2006ce, 
Feldman:2006wb,
Cheung:2007ut,  
Feldman:2007wj, 
Feldman:2009wv}. 
The Lagrangian for the extension is given by 
${\cal L} = {\cal L}_F + {\cal L}_W$ 
where 
\bea
&-4 {\cal L}_{F} = 
X_{\mu\nu}^2 +2 ( \partial_\mu \sigma_1 +  m_{1}\epsilon_1 B_{\mu} 
+  m_{1} X_{\mu} )^2, \nonumber \\ 
&-4{\cal L}_{W} = 
C_{\mu\nu}^2  +  2( \partial_\mu \sigma_2 +  m_{2}\epsilon_2 B_{\mu} 
+  m_{2} C_{\mu} )^2. \nonumber 
\eea
Here $B_\mu$ is the hypercharge boson in the SM, 
$\sigma_1$ and $\sigma_2$ are the axion fields 
{in the Stueckelberg mechanism},
and $m_1$, $m_2$, $m_{1}\epsilon_1$, 
and $m_{2}\epsilon_2$ are mass terms in 
the Stueckelberg mechanism.\footnote{Two additional mass parameters  
can be present in the Lagrangian so that 
$-4 {\cal L}_{F} = 
X_{\mu\nu}^2 +2 ( \partial_\mu \sigma_1 
+  m_{1}\epsilon_1 B_{\mu} 
+  m_{1} X_{\mu} 
+  m'_{1} C_{\mu} )^2$, 
and 
$-4{\cal L}_{W} = 
C_{\mu\nu}^2  +  2( \partial_\mu \sigma_2 +  m_{2}\epsilon_2 B_{\mu} 
+  m_{2} C_{\mu} +  m'_{2} X_{\mu} )^2$. 
In our analysis we take both $m'_{1}$ and $m'_{2}$ 
to be negligible for simplicity.}
{The dimensionless parameters 
$\epsilon_1$ and $\epsilon_2$ are assumed to be 
small in our analysis: 
we set $\epsilon_1 \sim {\cal O}(10^{-7})$ 
so that an LLDP at the LHC can arise; 
we set ${\epsilon_2} \sim {\cal O}(10^{-2})$ 
in order to obtain a significant LHC production cross section 
while satisfying various experimental constraints.}
${\cal L}_{F}$ is invariant under 
$U(1)_Y$ gauge transformations 
$
\delta_Y B_{\mu} = \partial_\mu  \lambda_B
$
and 
$
\delta_Y \sigma_1 = - m_{1}\epsilon_1  \lambda_B
$; 
${\cal L}_{F}$
is also invariant under $U(1)_F$ gauge transformation 
$
\delta_F X_{\mu} = \partial_\mu  \lambda_X ,
$
and 
$ 
\delta_F \sigma_1 = - m_{1} \lambda_X
$. 
Similarly, ${\cal L}_{W}$ is gauge invariant 
under both $U(1)_{Y}$ and $U(1)_{W}$.

In the hidden sector, we further introduce one Dirac fermion $\psi$ 
that is
charged under both  $U(1)_{F}$ and $U(1)_{W}$.
Vector current interactions between the 
Dirac fermion and the gauge bosons in the hidden sector 
are assumed, i.e., 
$
g_F \bar \psi \gamma^\mu \psi X_{\mu} 
+ g_W \bar \psi \gamma^\mu \psi C_{\mu} 
$
where $g_{F}$ and $g_W$ are the gauge couplings 
for $U(1)_{F}$ and $U(1)_{W}$ respectively.

The two by two mass matrix in the neutral gauge boson 
sector in the SM is now extended to a four by four mass matrix, 
which, in the gauge basis $V= ( C,X, B, A^3)$, 
is given by
\be
M^2 = 
\begin{pmatrix} 
m_{2}^2  & 0 & m_{2}^2 \epsilon_2 & 0\cr
0 & m_{1}^2 & m_{1}^2 \epsilon_1 & 0 \cr
m_{2}^2 \epsilon_2 & m_{1}^2 \epsilon_1 &  
\sum\limits_{i=1}^2 m_{i}^2 \epsilon_i^2 + {g'^2 v^2 \over 4} 
  & - {g'g v^2 \over 4}  \cr
0 &  0 & - {g'g v^2 \over 4} & {g^2 v^2 \over 4}
\end{pmatrix} 
\label{2zpstmass}
\ee
where $g$ and $g'$ are gauge couplings for 
the SM $SU(2)_L$ and $U(1)_Y$ 
gauge groups respectively,  
$A^3$ is the third component of the $SU(2)_L$ gauge 
bosons, 
and $v$ is the Higgs vacuum expectation value. 
The determinant of the mass square matrix 
vanishes which ensures the existence of a  
massless mode to be identified as the SM photon.

The mass matrix can be diagonalized via 
an orthogonal transformation
${\cal O}$ such that 
${\cal O}^T M^2 {\cal O} = 
{\rm diag} (m^2_{Z'}, m^2_{A'}, m^2_{Z}, 0)$; 
the mass {basis}
$E= ( Z', A', Z, A)$ {is}
related to the gauge {basis} $V$ via 
$E_i={\cal O}_{ji} V_{j}$. 
In the mass basis, 
$A$ is the SM photon, 
$Z$ is the SM $Z$ boson, 
$A'$ is the dark photon 
with {a GeV-scale} mass and $Z'$ is the 
{heavy boson with a TeV-scale mass.}
Diagonalization of the mass matrix leads to 
interactions between $Z/A$ in the SM 
to the fermion $\psi$ in the hidden sector, 
and also interactions between $Z'/A'$ in the hidden 
sector and SM fermions $f$; 
both $\psi-Z/A$ and $f-Z'/A'$ couplings are 
suppressed by the small $\epsilon_1$ and 
$\epsilon_2$ parameters, 
{and vanish in the $\epsilon_1=0=\epsilon_2$ limit.}

We parameterize the interactions between the fermions 
and the mass eigenstates of the neutral bosons via 
\be
\bar f \gamma_\mu (v^f_i - \gamma_5 a^f_i) f E^\mu_i 
+ v^\psi_i \bar \psi \gamma_\mu  \psi E^\mu_i 
\ee
where the vector and axial-vector couplings are given by 
\bea
\label{eq:vcouplings}
v^f_i &=& (g {\cal O}_{4i} - g' {\cal O}_{3i})T^{3 }_f/2
+ g' {\cal O}_{3i}Q_f , \\ 
\label{eq:acouplings}
a^f_i &=& (g {\cal O}_{4i} - g' {\cal O}_{3i})T^{3 }_f/2, \\
\label{eq:couplings} 
v^\psi_i &=& g_{W} {\cal O}_{1i} + g_{F} {\cal O}_{2i},
\eea
with $Q_f$ is the electric charge of fermion $f$, 
$T_f^3$ is the quantum number of the left-hand 
chiral component of the fermion $f$ 
{under SU(2)$_L$}.

{The eigenvector corresponding to the photon, the massless eigen mode, 
is ${\cal O}_{j4}=N(-\epsilon_2, -\epsilon_1,1, g'/g)$ where $N$ is the normalization. 
The SM fermion couplings to the photon are thus given by 
$v^f_4 = e Q_f$ and $a^f_4 = 0$ 
where the coupling constant $e=g'N$. 
As in SM, the photon here couples to the SM fermions 
only via the vector coupling that is proportional to the electric charge 
of the fermion. 
The coupling constant $e$ is now related to the gauge 
couplings $g$ and $g'$ via 
\be
\frac{1}{e^2} = \frac{1}{g^2} + \frac{1+\epsilon_1^2 + \epsilon_2^2}{g'^2} 
\equiv  \frac{1}{g^2} + \frac{1}{g_{\rm SM}^{'2}} 
\ee
where $g'_{\rm SM}$ is the effective coupling constant in the SM. 
The hidden fermion $\psi$ interacts with the SM photon 
via the vector coupling 
$
v^\psi_4 =  -(e/g') (\,  \epsilon_2\, g_{W}  + \,\epsilon_1 \, g_{F} ) 
\equiv e \delta
$
where $\delta$ is a very small electric charge 
which is often referred to as ``millicharge'' 
in the literature. 
We also present approximated 
expressions of the vector and axial-vector 
couplings for the three massive eigen modes 
in Appendix \ref{app:couplings}. 
}
%
%


\section{Experimental constraints}
\label{sec:constraints}

Here we discuss various constraints on the model, 
including electroweak constraints from LEP, 
constraints from LHC, and also cosmological constraints.

In our analysis, we will fix most model parameters so that 
a sizable LHC signal is expected. We first discuss these 
default parameter values.
The heavy $Z'$ boson in our model mostly originates from the 
$U(1)_W$ boson $C_\mu$; 
{we choose $m_2 = 700$ GeV as the benchmark point 
in our analysis.}\footnote{The LLDP signal at the LHC  
is insensitive to the $m_2$ value 
when $m_2 > {\cal O} (100)$ GeV.}
In order to obtain a sufficient large $\psi\bar\psi$ 
production cross section at the LHC, we choose $g_W = 1$; 
a relatively large $U(1)_F$ coupling constant 
is also chosen, $g_F = 1.5$, 
so that a rather sizable dark radiation rate 
for the $\psi$ particle can be achieved in the model. 
\footnote{We set $g_W = 1$ and $g_F = 1.5$ 
at the $A'$ mass scale. The 
RGE running of $g_W$ is insignificant, 
$\lesssim $ 5\%  increment on $g_W$ at  
the $Z'$ mass scales.}
We use 
$c\tau = \hbar c /\Gamma = 1$ m 
as the characteristic value for 
the proper lifetime of the dark photon, 
where $\Gamma$ is the dark photon decay width; 
the dark photon decay widths are given in Appendix \ref{app:decay}. 
We find that small modifications around 
$c\tau = 1$ m do not lead to significant changes in 
the collider signatures.
The above values are the default ones used 
throughout the analysis, if not explicitly specified.

\begin{figure}[htbp]
\begin{centering}
\includegraphics[width=0.5\textwidth]{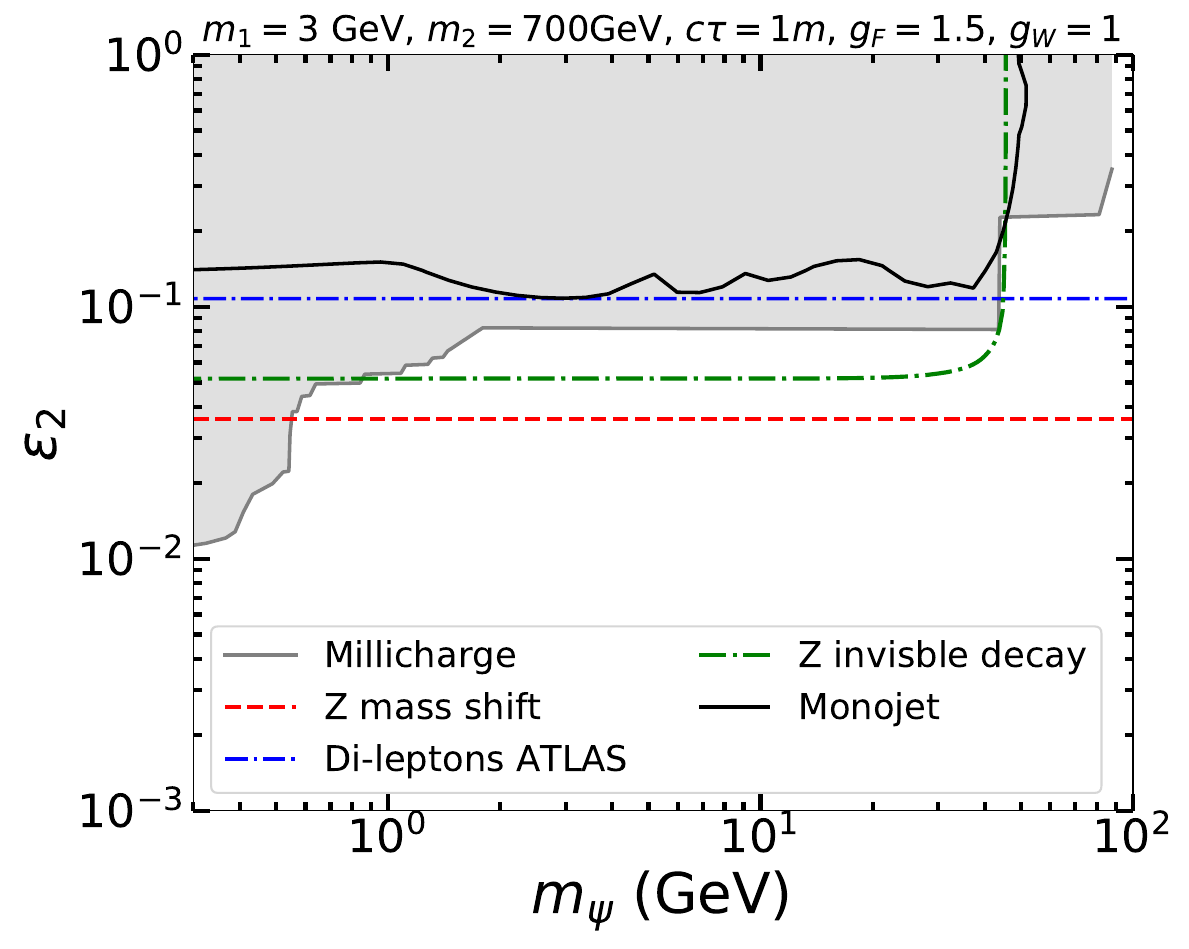}
\caption{The $95\%$ CL exclusion limits on 
$\epsilon_2$ 
as a function of $m_\psi$. 
We use the following parameters: 
$m_1 =  3$ GeV, $m_2 =  700$ GeV, 
$g_F = 1.5$, $g_W = 1.0$, 
and $c\tau = 1$ m. 
Shown are 
the millicharged particle searches at the colliders (shaded light gray) 
\cite{Davidson:2000hf},
electroweak constraints due to  
the $Z$ mass shift (dashed red), 
the $Z$ invisible decay (dash-dotted green) \cite{ALEPH:2005ab}, 
the di-lepton high mass resonance search 
at ATLAS (dash-dotted blue)  \cite{ATLAS:2019vcr}, 
and the mono jet search at ATLAS (solid black) 
\cite{Aaboud:2017phn}.
}
\label{fig:limit-mchi-eps2} 
\end{centering}
\end{figure}

{\it $Z$ invisible decay:} 
The $Z$ invisible decay width is measured to be  
$\Gamma_{\rm inv}^{\rm Z} \pm \delta \Gamma^{\rm Z}_{\rm inv} 
= 499$ MeV $\pm 1.5$ MeV \cite{ALEPH:2005ab}. 
The $Z$ boson can decay into the 
$\psi\bar\psi$ final state, if $m_Z > 2 m_\psi$, 
with a decay width 
\be
\Gamma_{Z \to \psi \bar{\psi}}=
{ m_{Z}  \over 12 \pi} (v^{\psi}_3)^{2}
\sqrt{1-4 x_{\psi Z}} (1+2 x_{\psi Z}),
\ee
where $x_{\psi Z} \equiv (m_{\psi}/m_{Z})^{2}$, 
and $v^{\psi}_3$ is the vector coupling between the $Z$ boson 
and $\psi$, as given in Eq. ({\ref{eq:couplings}}). 
Equating the invisible decay width due to the $\psi\bar\psi$ final state 
to the experimental uncertainty $\delta \Gamma^{\rm Z}_{\rm inv}$,  
one obtains an upper bound on $v^{\psi}_3$, 
which is shown on Fig.\ ({\ref{fig:limit-mchi-eps2}}). 
For light $\psi$ mass, one has 
$v^{\psi}_3 \gtrsim 2.5 \times 10^{-2}$. 

%

{\it Electroweak constraint on the $Z$ mass: }
The mass of the $Z$ boson is modified due to the 
enlarged neutral gauge boson mass matrix, 
as given in Eq.\ (\ref{2zpstmass}). 
For the parameter space of interest in our analysis, 
i.e., $\epsilon_1 \ll \epsilon_2$, 
the mass shift on the $Z$ boson can be estimated as 
\be
\left| {\Delta m_{Z} \over m_{Z}}\right|
\simeq {\epsilon_2^{2} \over 2} s^2_W
\left(1- {m_{Z}^{2} \over m_{2}^{2}} \right)^{-1}, 
\label{MZdev}
\ee
where $s^2_W \equiv \sin ^{2} \theta_{W} = 0.22343$ \cite{Tanabashi:2018oca}, 
with $\theta_{W}$ being the weak rotation angle.
We adopt the methodology in Ref.\ \cite{Feldman:2006ce} 
to estimate the electroweak constraints. 
The experimental uncertainty of the $Z$ mass is given by \cite{Feldman:2006ce} 
\be
\bigg[ {\delta m_Z \over m_Z} \bigg]^{2}=
\bigg[ {c_W^{-2}-2 t_W^2 \over \delta m_{W}^{-1}  m_W}\bigg]^{2}
+\frac{t_W^4(\delta \Delta r)^{2}}{4(1-\Delta r)^{2}},
\label{deltaMZ}
\ee
where 
$c_W \equiv \cos\theta_{W}$, 
and $t_W \equiv \tan\theta_{W}$.
Here we take into account the recent analysis on the 
uncertainty of the $W$ boson mass , 
$m_W \pm \delta m_W = 80.387 \pm 0.016$ GeV \cite{Tanabashi:2018oca}, 
and on the radiative correction, 
$\Delta r \pm \delta \Delta r =  0.03672 \mp 0.00017 \pm 0.00008$ \cite{Tanabashi:2018oca} 
where the first uncertainty is due to the top quark mass 
and the second is due to the fine structure constant $\alpha({m_Z)}$ 
at the $m_Z$ scale. 
Adding in quadrature, we obtain $\delta \Delta r =  0.00019$. 
Equating the mass shift on the $Z$ boson given in 
Eq.\ (\ref{MZdev}) to the experimental uncertainty, 
one obtains an upper bound on $\epsilon_2$ 
\be
|\epsilon_2| \lesssim 0.036  \,\sqrt{1-\left(m_{Z} / m_{2}\right)^{2}}.
\label{eq:eps2MZ}
\ee
The upper limit analyzed in this section using only the $Z$ mass constraint  is similar to
that from a global-fit analysis on a number of Z-pole observables \cite{Feldman:2006ce}. 
Using the $Z$ mass constraint, we can also reproduce the limit obtained in a more recent 
electroweak global analysis \cite{Huang:2019obt}.  
Thus the method using only the $Z$ mass constraint provides a quick way 
to estimate the electroweak constraints. 
%

{\it  Di-lepton constraint on $Z'$ decays: }
{Dilepton final states which are produced 
at the LHC from the heavy $Z'$ boson 
via the Drell-Yan process, can be searched 
for by reconstructing their invariant mass.}
Fig.\ ({\ref{fig:limit-mchi-eps2}}) shows the 
95 \% CL upper bound on $\epsilon_2$ 
in the dilepton channel from ATLAS \cite{ATLAS:2019vcr}.

\begin{figure}[htbp]
\begin{centering}
\includegraphics[width=0.5\textwidth]{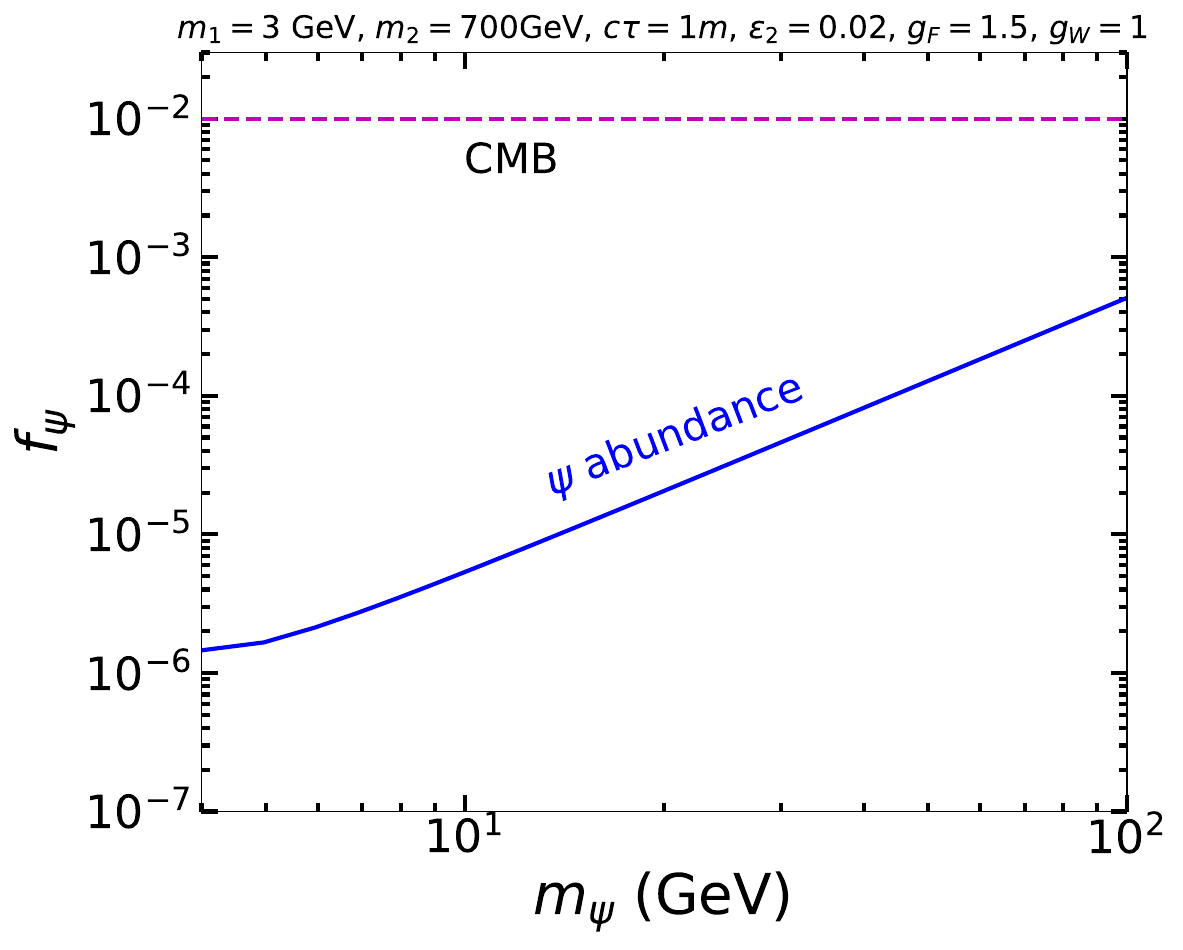}
\caption{The fraction of $\psi$ to the total DM in the Universe 
as a function of the $\psi$ mass. 
Here, we take the canonical DM cross section is $\langle \sigma v\rangle_{\rm DM} = 1$ pb.
The magenta dashed line represents the current limit from 
Refs. \cite{Munoz:2018pzp, Boddy:2018wzy, dePutter:2018xte, Kovetz:2018zan}. 
}
\label{fig:relic} 
\end{centering}
\end{figure}

{\it Constraints on millicharge}: 
In the parameter space of interest of the model, 
the hidden fermion $\psi$ possesses a millicharge 
$\delta \simeq \epsilon_2\, g_W/g'$.
The $\psi$ particle remains undetectable in 
typical particle detectors due to 
the minute electric charge. 
Fig.\ (\ref{fig:limit-mchi-eps2}) shows the collider constraints on 
millicharge in the ($m_\psi$, $\epsilon_2$) plane 
\cite{Davidson:2000hf}. 
The collider constraints on millicharge are 
the most stringent constraints on $\epsilon_2$
for $m_\psi \lesssim 0.6$ GeV.

Because the $\psi$ particle is charged under 
$U(1)_W$ and $U(1)_F$ in the hidden sector, it is stable 
and thus can be a dark matter (DM) candidate. 
For the parameter space of interest, i.e., $m_\psi>m_{A'}$, 
the annihilation channel into on-shell dark photons, 
$\psi \bar{\psi} \to A' A'$, is the dominant process  
for the relic abundance of the $\psi$ particle; 
the annihilation cross section can be approximated as follows
\cite{Cline:2014dwa}
\be
\langle \sigma v\rangle_{\psi \bar{\psi} \to A' A'} 
\simeq \frac{(v^\psi_2)^4}{16 \pi \,m_{\psi}^2} \frac{(1-r^2)^{3/2}}{(1-r^2/2)^{2}},
\label{eq:chichiz1z1}
\ee
where $v^\psi_2$ is the coupling between the dark photon and $\psi$, 
as given in Eq.~(\ref{eq:couplings}), and $r = m_{A'}/m_{\psi}$. 
We compute the ratio between the $\psi$ relic abundance 
and the total DM relic abundance via  
$f_{\psi} = 2 \langle \sigma v\rangle_{\rm DM}/\langle \sigma v\rangle_{\psi \bar{\psi} \to A' A'}$, 
where $\langle \sigma v\rangle_{\rm DM} = 1$ pb is the canonical DM cross section, 
and the factor of 2 accounts for the Dirac nature of the $\psi$ particle. 
Because $v^\psi_2\simeq g_F$, the annihilation cross section 
given in Eq.~(\ref{eq:chichiz1z1}) 
is much larger than the canonical annihilation cross section needed 
for the cold DM relic density in the Universe 
\cite{Ade:2015xua}, 
for the case $m_{A'} \sim {\cal O}$(1) GeV and $m_\psi \sim {\cal O}$(10) GeV. 
Thus, the contribution of the $\psi$ particle to the DM in the Universe 
is less 0.1\% when $m_\psi < 100$ GeV, as shown in Fig.\ (\ref{fig:relic}). 
This is consistent with the cosmological limits on millicharged DM, 
which constrain the fraction of the millicharged DM to be 
$ \lesssim 1\%$ of the total DM in the Universe 
\cite{Munoz:2018pzp, Boddy:2018wzy, dePutter:2018xte, Kovetz:2018zan}.
The $\psi$ DM is efficiently stopped by the rock above underground labs 
of DM direct detection experiments,  
unless the millicharge is extremely small.
Adapting the estimation in Refs. \cite{Foot:2003iv,Cline:2012is} for 1 km of rock, 
we found that the DM direct detection is only sensitive 
to the mixing parameter of $\epsilon_{2} < 10^{-6}$ in our model. 
Thus the current underground DM direct detection experiments do not 
constrain the model.

{\it Monojet constraints}: 
Searches for invisible particles which are 
produced in association with 
an initial state radiation (ISR) jet have been carried out 
at ATLAS \cite{Aaboud:2017phn} and CMS \cite{Sirunyan:2017jix}.
Here {we} recast the ATLAS result \cite{Aaboud:2017phn} 
to set constraints on our model. 
We use MadGraph5 aMC@NLO (MG5) \cite{Alwall:2014hca} 
to generate events for the process $p p \to \psi \bar{\psi} j $ 
which are then passed to Pythia 8 for  
showering and hadronization \cite{Sjostrand:2014zea, Carloni:2010tw, Carloni:2011kk}. 
The Madanalysis 5 package \cite{Dumont:2014tja, Sengupta} 
is further used to analyze
the ATLAS results \cite{Aaboud:2017phn}. 
We use the same detector cuts as in Ref.\ \cite{Aaboud:2017phn}; 
the optimal selection region for our model is found to be in 
the window: $E_{T}^{\rm miss} \in (300,350)$ GeV 
(the EM2 region in Ref.\ \cite{Aaboud:2017phn}). 
The $95\%$ {CL} exclusion limit 
on $\epsilon_2$ from the monojet channel in ATLAS 
\cite{Aaboud:2017phn} is shown in Fig.\ (\ref{fig:limit-mchi-eps2}). 
When $m_{\psi} < m_Z/2$, the $Z$ boson diagram, 
i.e., the $p p \to Z j \to \psi \bar{\psi} j$ process gives 
the dominant contribution to the monojet signal; 
when $m_{\psi} > m_Z/2$, the $Z'$ boson diagram, 
i.e., the $p p \to Z' j \to \psi \bar{\psi} j$ becomes 
more important than the $Z$ process. 
Because of the large $Z'$ mass, 
the $Z'$ boson diagram is suppressed 
as compared to the $Z$ diagram. 
Thus the monojet channel only provides a comparable  
constraint to other constraints 
for $m_{\psi} < m_Z/2$. 

As shown in Fig.\ (\ref{fig:limit-mchi-eps2}), 
the electroweak constraint on the $Z$ mass shift 
provides the best constraint to the parameter space 
of interest in our model, 
except for the $m_\psi \lesssim 0.6$ GeV region 
where collider constraints on millicharged particles 
become strong.

\section{Timing detector}

Recently, some precision timing detectors are proposed to be installed 
at CMS \cite{CMStiming}, ATLAS \cite{Allaire:2018bof, Allaire:2019ioj} and LHCb \cite{LHCbtiming}.
These timing detectors, which aim to reduce the pile-up rate at 
{the high luminosity LHC (HL-LHC)}, 
can also be used in long-lived particle searches 
\cite{Liu:2018wte, Mason:2019okp,Kang:2019ukr, Cerri:2018rkm}. 
In this analysis, we focus on one of the timing detectors, 
the minimum ionizing particle (MIP) timing detector 
to be installed to the CMS detector (hereafter MTD CMS) \cite{CMStiming}, 
which has a $\sim$30 pico-second timing resolution.
At the MTD CMS, the timing 
detection layer, which 
is proposed to be installed between 
the inner tracker and the electromagnetic calorimeter, 
is about  
1.17 m away from the beam axis and 6.08 m 
long in the beam axis direction. 
A time delay at the LHC due to 
long-lived particles can be measured by the 
new precision timing detectors, 
which can enhance the 
collider sensitivity to such models \cite{Liu:2018wte}.

\begin{figure}[htbp]
\begin{centering}
\includegraphics[width=0.35\textwidth]{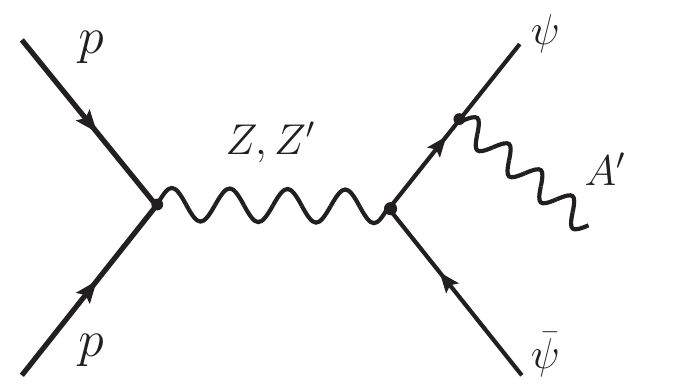} 
\caption{Feynman diagram for the dark photon production at the LHC.}
\label{fig:feyndiag}
\end{centering}
\end{figure}

At the LHC, the hidden sector particle $\psi$ 
can be pair-produced via $pp \to Z/Z' \to \bar \psi \psi$, 
which subsequently 
radiates dark photons 
$\psi \to \psi A'$; 
the corresponding Feynman diagram is shown 
in Fig.\ (\ref{fig:feyndiag}).
Due to the feeble interaction strength to SM fermions, 
the $A'$ boson travels a 
macroscopic distance away from its production point 
and then decays into a pair of SM particles
which are detected by 
the timing layers. 
Here we use the di-lepton final states measured by 
the timing layers to detect the LLDP in our model. 
The time delay between the leptons from   
the LLDP and SM 
particles produced at the primary vertex 
is given by 
\be
\Delta t = {L_{A'}/v_{A'}} + 
L_\ell - L_{\rm SM},
\ee
where the $L$'s are the distances traveled by 
various particles and $v=c$ for the SM particles \cite{Liu:2018wte}. 
The time delay is significant if the LLDP 
moves non-relativistically.

We select the leading leptons with transverse momentum 
$p_{T}^{\ell} > 3$ GeV to suppress 
faked signals from hadrons produced with low $p_T$ 
\cite{Sirunyan:2017zmn} and to ensure 
that the lepton is moving relativistically and 
its trajectory is not significantly bent \cite{Liu:2018wte}. 
The point where the dark photon decays 
is required to have 
a radial distance away from the beam axis of  
$0.2 ~{\rm m} < L^T_{A'} < 1.17 ~{\rm m}$ and 
a longitudinal distance along the beam axis  
of $|z_{A'}| < 3.04  ~{\rm m}$.  
Following Ref. \cite{Liu:2018wte}, 
an ISR jet with $p_T^j > 30$ GeV and $|\eta_j| <2.5$ 
is required to {time stamp} the hard collision. 
The time delay is required to be $\Delta t > 1.2 $ ns 
in order to suppress the background.

The dominant SM backgrounds 
come from the multi trackless jets in the same-vertex (SV) 
hard collisions 
and in the pile-up (PU) events \cite{Liu:2018wte}, 
as well as photons in SV hard collisions \cite{Mason:2019okp}. 
The SV background arises because of 
the finite timing resolution; 
the PU background is due to the fact that within one bunch crossing, 
two hard collisions occurring at two different times 
can lead to a time delay signal.

We compute the dijet events at the LHC with $\sqrt{s}=13$ TeV 
by using MG5 \cite{Alwall:2014hca} and also Pythia 8 \cite{Sjostrand:2014zea}. 
We select events in which the leading jet has 
$p_T > 30$ GeV and $|\eta(j)| < 2.5$ to time stamp the 
primary collision, and the subleading jets have 
$p_T^j > 3$ GeV and $|\eta(j)| < 2.5$. 
The inclusive jet cross section is 
$\sigma_j \approx 1\times 10^{8}$ pb, 
under these detector cuts.

The inclusive photon production cross section at NLO is 
$\sigma_{\gamma} \approx 2 \times 10^{8}$ pb 
at the LHC with $\sqrt{s}=13$ TeV, 
by using JETPHOX \cite{Catani:2002ny} 
with the CT10 PDF. 
The detector cuts are  
$p_T^{\gamma} > 3$ GeV and $|\eta^{\gamma}| < 2.5$ for photon 
and $p_T^{j} > 30$ GeV and $|\eta^{j}| < 2.5$ for the leading jet.

At the 13 TeV LHC, the SV background events 
can be estimated as \cite{Liu:2018wte, Mason:2019okp} 
\be 
N_{\rm SV} = \sigma_{\gamma}{\cal L} + \sigma_{j}{\cal L} f_{\gamma}
\sim 6\times 10^{14},
\ee 
where ${\cal L} = 3{\rm \, ab^{-1}}$ is the integrated luminosity, 
and $f_{\gamma} \approx 10^{-4}$ is the rate of a jet to fake a photon or a lepton 
\cite{Mason:2019okp}.
The PU background events can be estimated as \cite{Liu:2018wte, Mason:2019okp}
\be
N_{\rm PU} = \sigma_{j} {\cal L} (n_{\rm PU} \frac{\sigma'_{j} }
{\sigma_{\rm inc}}) f_{\gamma} f_{j} \sim 3.75\times10^{9},
\ee 
where $f_{j} \sim 10^{-3}$ \cite{Mason:2019okp} is the rate for the jet to be trackless, 
$\sigma'_{j} \approx 1\times 10^{11}$ pb is 
the dijet cross section with the requirement on all jets of $p_T^j > 3$ GeV and $|\eta(j)| < 2.5$, 
$\sigma_{\rm inc} = 80$ mb \cite{Aaboud:2016mmw} 
is the inelastic cross section of $pp$ collisions at 13 TeV 
and $n_{\rm PU} \approx 100$ \cite{Sopczak:2017mvr} 
is the average pile-up number at the HL-LHC.

The time delay distribution of the SM background 
can be described by a Gaussian distribution \cite{Liu:2018wte} 
\be
\frac{d{\cal P}(\Delta t)}{d\Delta t} = \frac{1}{\sqrt{2\pi} \delta_t} e^{\frac{-\Delta t^2}{2\delta_t^2}},
\ee 
where $\delta_t$ is the time spread. 
For the PU background, the time spread, $\delta_t = 190$ ps, 
is determined by the beam property; 
for the SV background, $\delta_t = 30$ ps,  
is determined by the time resolution \cite{CMStiming}. 
We find that under the detector cut 
$\Delta t > 1$ ns, 
the SV background is negligible 
and the PU background is about 260 with ${\cal L} = 3{\rm \, ab^{-1}}$; 
the PU background also becomes negligible, 
$N_{\rm PU} \lesssim 0.5$, 
if the time delay $\Delta t > 1.2$ ns is required. 
Thus, we take $\Delta t > 1.2$ ns as the detector cut 
in our analysis.

\begin{figure}[htbp]
\begin{centering}
\includegraphics[width=0.5\textwidth]{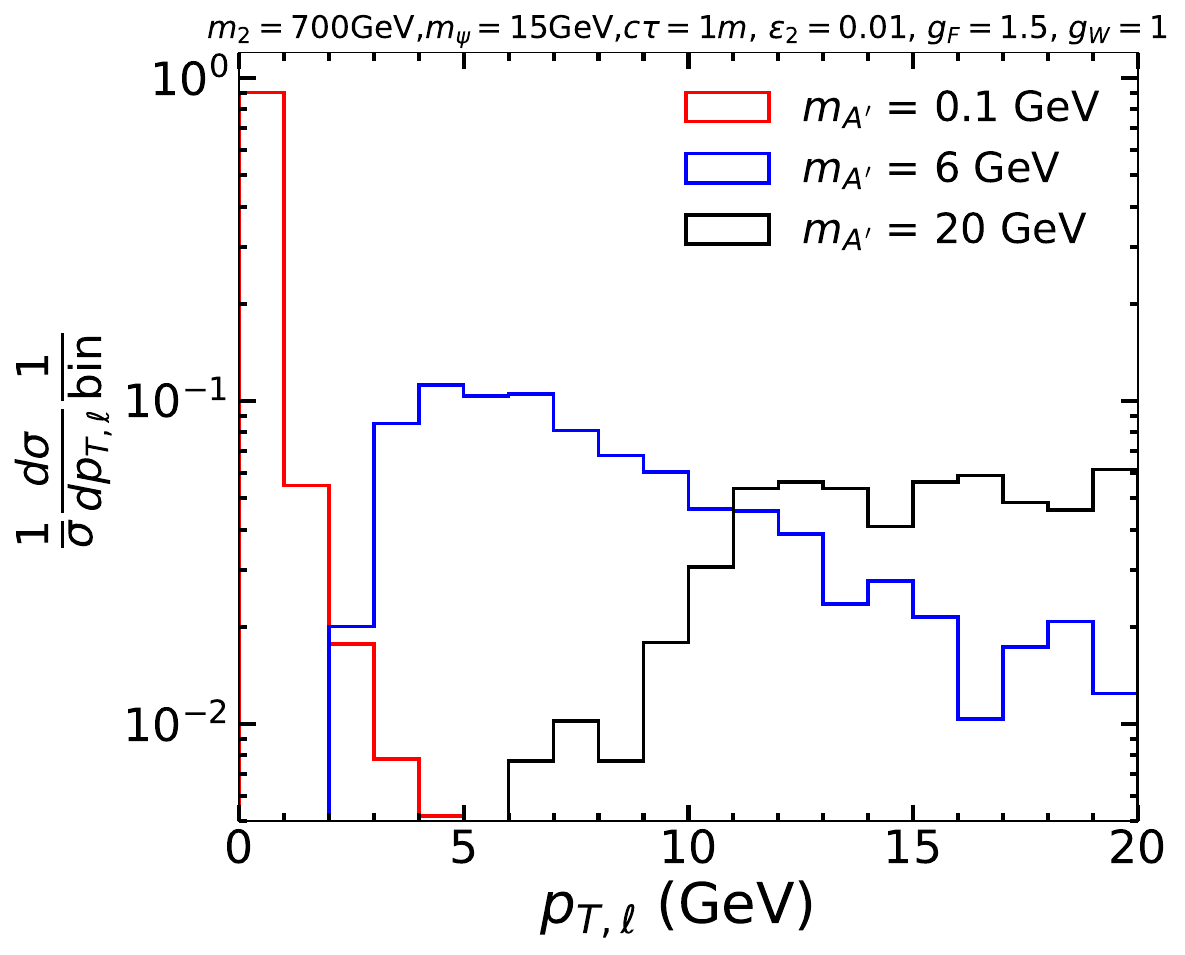}
\caption{
The $p_T$ distribution of the leading lepton. 
We choose $m_2 =  700$ GeV, 
$m_{\psi} =  15$ GeV, $c\tau = 1$ m, $\epsilon_2 = 0.01$, $g_F = 1.5$ and $g_W = 1.0$. 
The red, blue and black lines 
indicate the dark photon mass of 0.1 GeV, 6 GeV and 20 GeV respectively.  
}
\label{fig:pTl-final}
\end{centering}
\end{figure}

\begin{figure}[htbp]
\begin{centering}
\includegraphics[width=0.5\textwidth]{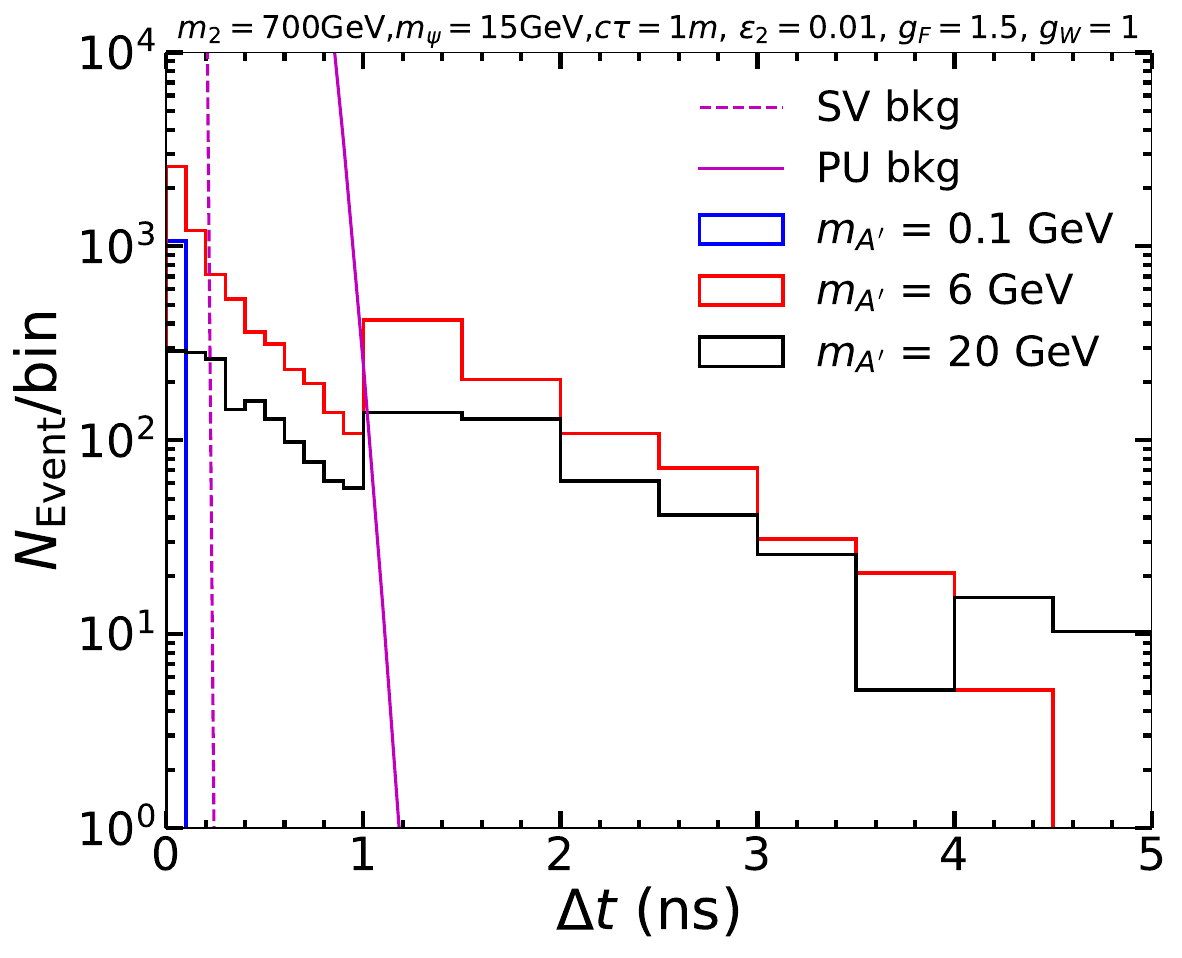}
\caption{The distribution of the time delay $\Delta t$, 
with the integrated luminosity ${\cal L} = 3 {\rm ab}^{-1}$.
$p_{T} > 3$ GeV is required for the leading lepton.
The bin width is 0.5 (0.1) ns for $\Delta t > 1$ ns ($\Delta t < 1$ ns). 
We choose
$m_2 =  700$ GeV, $m_{\psi} =  15$ GeV, 
$c\tau = 1$ m, $\epsilon_2 = 0.01$, $g_F = 1.5$ and $g_W = 1.0$. 
The solid blue, red and black lines 
indicate the dark photon mass of 0.1 GeV, 6 GeV and 20 GeV.  
The solid and dashed magenta curves represent the PU and SV backgrounds 
respectively.}
\label{fig:delt-final}
\end{centering}
\end{figure}

\begin{figure}[htbp]
\begin{centering}
\includegraphics[width=0.5\textwidth]{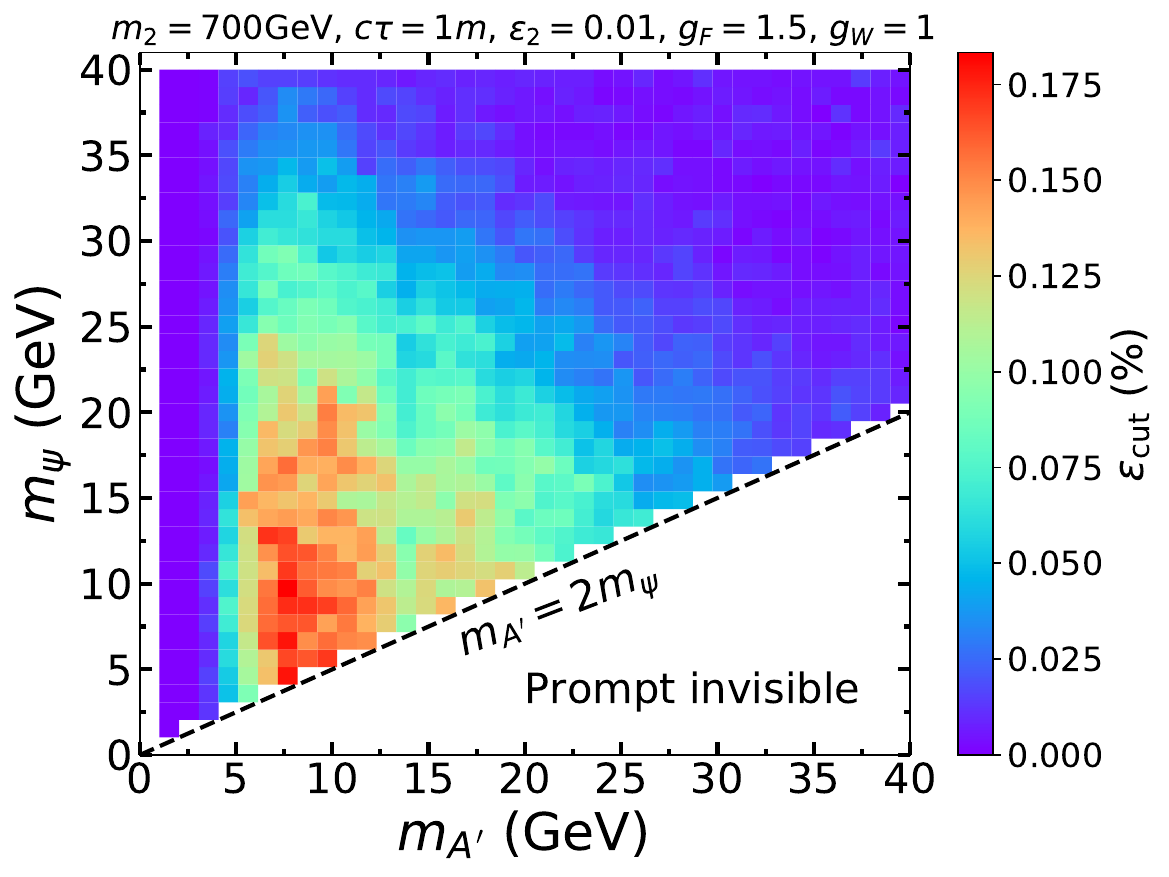} 
\caption{The cut efficiency $\epsilon_{\rm cut}$ as a function of $m_{A'}$ and $m_\psi$. 
We set $m_2 =  700$ GeV, $c\tau = 1$ m, $\epsilon_2 = 0.01$, $g_F = 1.5$ and $g_W = 1.0$. 
}
\label{fig:eff}
\end{centering}
\end{figure}

\begin{figure}[htbp]
\begin{centering}
\includegraphics[width=0.5\textwidth]{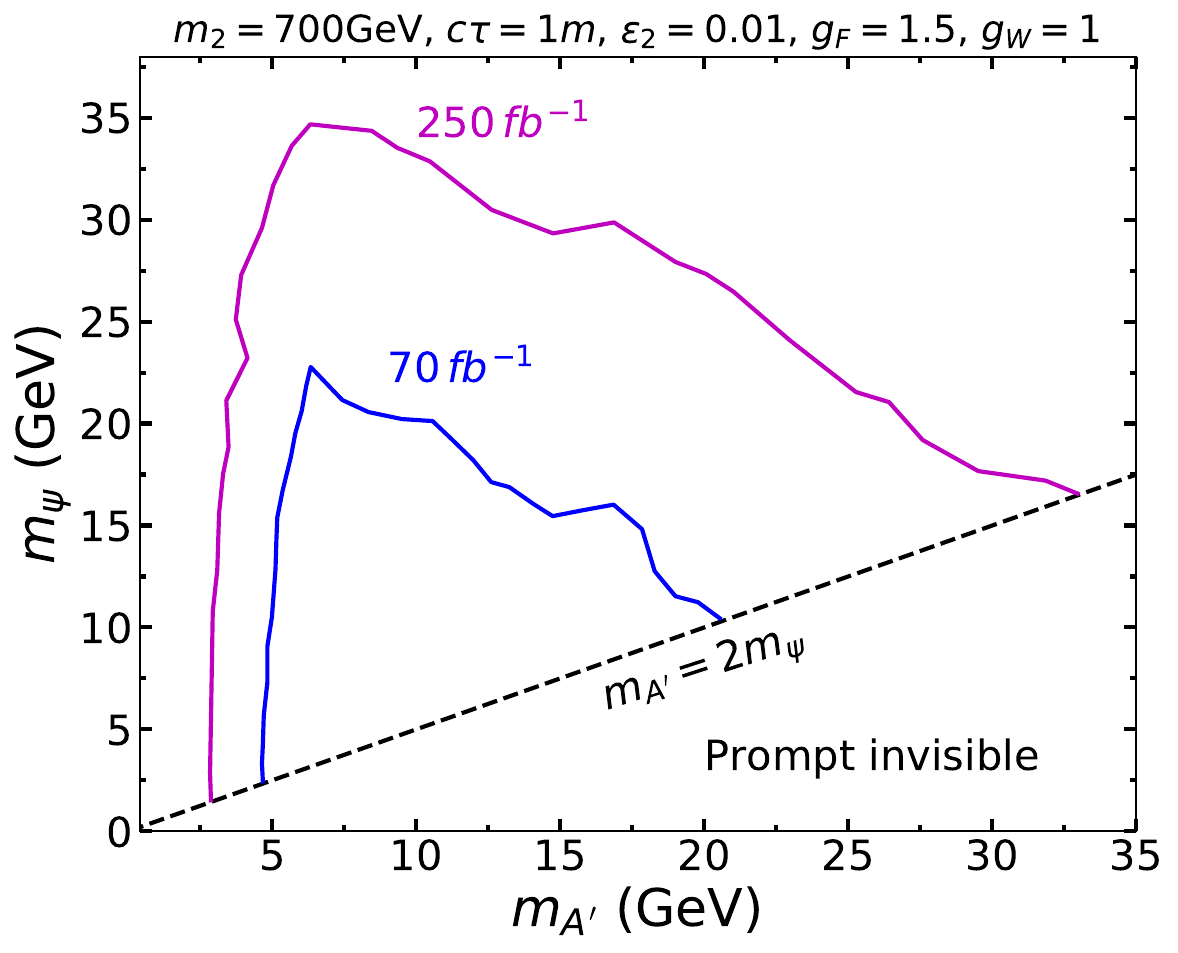} 
\caption{The contour of the expected signal events 
at the 13 TeV LHC, as a function of $m_{A'}$ and $m_\psi$. 
We choose $m_2 =  700$ GeV, $c\tau = 1$ m, $\epsilon_2 = 0.01$, $g_F = 1.5$ and $g_W = 1.0$. 
The blue and magenta contours indicate the needed integrated luminosity of 
70 fb$^{-1}$ and 250 fb$^{-1}$ respectively, to generate 10 events. 
}
\label{fig:limit}
\end{centering}
\end{figure}

We perform a full MC simulation and study the efficiency of the detector cuts 
in the parameter space of interest. 
We first implement the model into {the} FeynRules package \cite{Alloul:2013bka} 
and pass the UFO model file into MG5 \cite{Alwall:2014hca} 
to generate $8\times 10^4$ events of the $\psi$ pair production associated 
{with the time stamping ISR jet}
i.e $p p \to \psi \bar{\psi} j$.  
The dark showering is simulated in  Pythia 8 
\cite{Sjostrand:2014zea, Carloni:2010tw, Carloni:2011kk}.

Fig.~(\ref{fig:pTl-final}) shows 
the transverse momentum distribution of the leading lepton, 
for three different dark photon masses. 
We choose $m_2 =  700$ GeV, $m_{\psi} =  15$ GeV, 
$c\tau = 1$ m, $\epsilon_2 = 0.01$, $g_F = 1.5$ and $g_W = 1.0$, 
as the benchmark point. 
The final state leptons from dark photon decays are generally 
not very energetic in the models shown in Fig.~(\ref{fig:pTl-final}). 
In particular, the lepton events are highly suppressed 
under the detector cut $p_T > 3$ GeV, 
for the $0.1$ GeV dark photon case. 

Fig.~(\ref{fig:delt-final}) shows the distribution of the time delay $\Delta t$. 
The model parameters in Fig.~(\ref{fig:delt-final})  are the same as in Fig.~(\ref{fig:pTl-final}). 
The SM backgrounds are negligible when the time delay $\Delta t > 1.2$ ns. 
When the dark photon becomes heavier, 
more events with a larger {time delay} appear, 
as shown {in} Fig.~(\ref{fig:delt-final}), 
since in this case, the dark photon has a higher 
probability to move non-relativistically. 
The increase of the events with 
the larger {time delay}, however, is offset by the smaller dark photon 
radiation rate of the heavier dark photon. 

Fig.~(\ref{fig:eff}) shows the cut efficiency as 
a function of $m_{A'}$ and $m_{\psi}$, 
where 1370 grid points are simulated. 
We set $m_2 =  700$ GeV, $c\tau = 1$ m, 
$\epsilon_2 = 0.01$, $g_F = 1.5$ 
and $g_W = 1.0$. 
As shown in Fig.~(\ref{fig:pTl-final}), 
the detector cut: $p_T>3$ GeV for the leading lepton, 
significantly reduces the efficiency for  
light dark photon mass. 
The low efficiency in the heavy mass region in Fig.~(\ref{fig:eff}) 
is primarily due to the low radiation rate \cite{Chen:2018uii}. 
It turns out that the region with significant cut efficiency 
has $5 \,{\rm GeV} < m_{A'}, m_\psi < 35 \,{\rm GeV}$, 
with the highest efficiency $\sim 0.18\%$. 
We note that, for the $m_{A'} > 2 m_{\psi}$ region, 
the dark photon is no longer a long-lived particle 
since it can decay into a pair of $\psi$. 

Fig.~(\ref{fig:limit}) shows the regions that 
can be probed at the 13 TeV LHC, with the 
discovery criterion: $S=10$,  
as a function of the dark photon mass and the $\psi$ mass. 
The number of signal events is computed via 
$S =\epsilon_{\rm cut}\,  {\cal L} \, \sigma(p p \to \psi \bar{\psi} j) $, 
where ${\cal L}$ is the integrated luminosity, 
$\sigma(p p \to \psi \bar{\psi} j)$ is the production cross section at the LHC, 
and $\epsilon_{\rm cut}$ is the cut efficiency as shown in Fig.~(\ref{fig:eff}).
The model parameters are the same as in Fig.~(\ref{fig:eff}).
The blue and magenta contours indicate 
the needed integrated luminosity to generate 10 signal events. 
Therefore, with an integrated luminosity of $70$ $\rm{fb}^{-1}$ at the HL-LHC, 
the LLDP can be discovered in the time delay channel 
in the mass region: $5$ GeV $< m_{A'},m_{\psi} < 21$ GeV, 
with the rest of the model parameters 
fixed as in Fig.~(\ref{fig:eff}). 
A larger mass region: 3 GeV$< m_{A'}, m_{\psi} <$ 30 GeV, 
can be discovered if $250$ $\rm{fb}^{-1}$ data are 
accumulated at the HL-LHC.

\begin{figure}[htbp]
\begin{centering}
\includegraphics[width=0.5\textwidth]{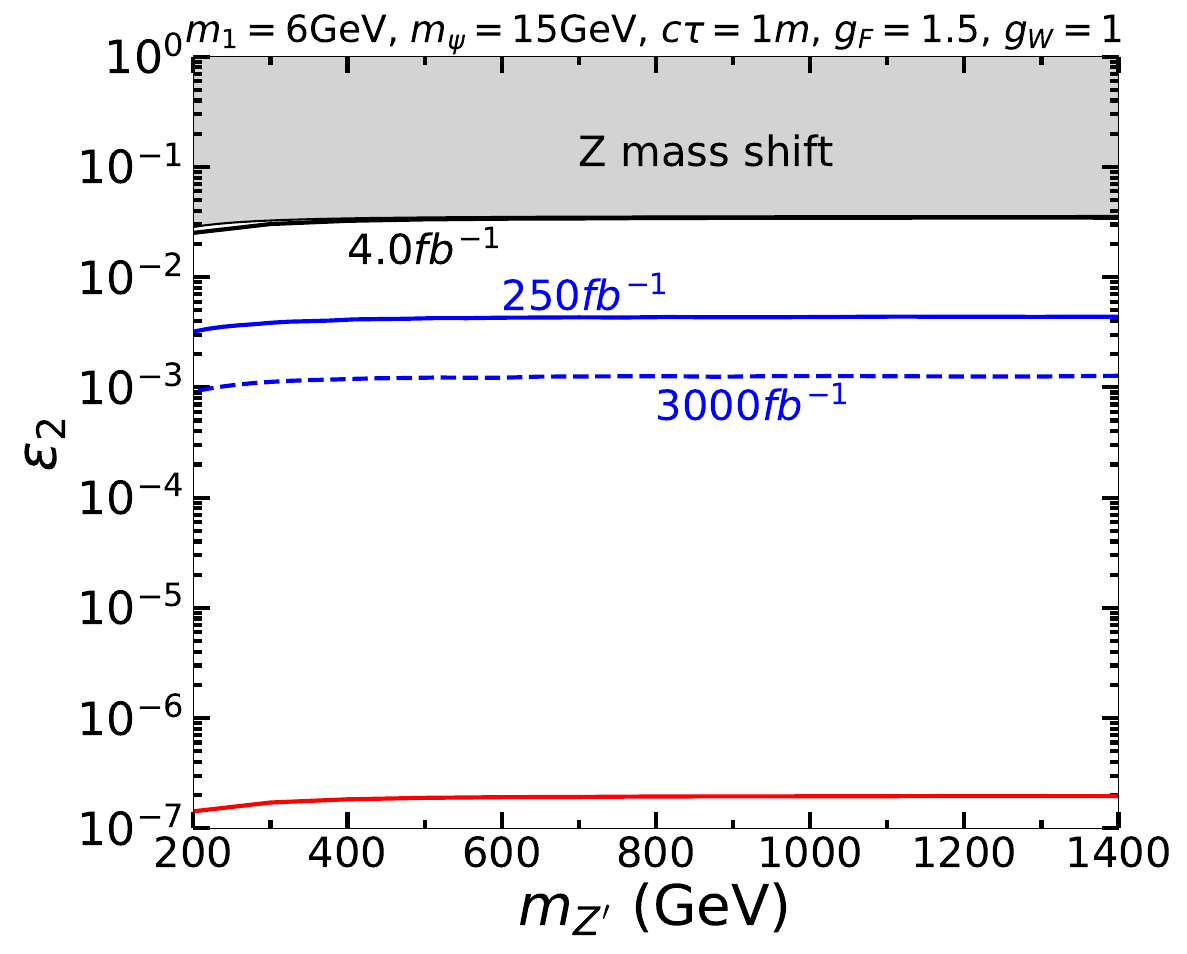} 
\caption{
The integrated luminosity needed at the HL-LHC to probe the parameter 
region spanned by $m_{Z'}$ and $\epsilon_2$. 
We choose $m_1 = 6$ GeV, $m_{\psi} = 15$ GeV, 
$c \tau = 1$ m, $g_F = 1.5$ and $g_W = 1$. 
The integrated luminosity needed are 
$\sim 4.0$ fb$^{-1}$ (black solid), 
$250$ fb$^{-1}$ (blue solid), 
and $3000$ fb$^{-1}$ (blue dashed).  
The gray shaded region is excluded by 
the $Z$ mass shift constraint,  
as given in Eq.\ (\ref{eq:eps2MZ}). 
Below the red line, the dark photon production cross section via 
$\epsilon_1$ dominates.}
\label{fig:signalregion}
\end{centering}
\end{figure}

Fig.\ (\ref{fig:signalregion}) shows the integrated 
luminosity needed to probe the parameter space 
spanned by $\epsilon_2$ and $m_{Z'}$. 
We choose $m_1 = 6$ GeV, $m_{\psi} = 15$ GeV, 
$c \tau = 1$ m, $g_F = 1.5$ and $g_W = 1$ as a benchmark. 
With an integrated luminosity of $\sim 4.0$ fb$^{-1}$, 
one  can probe the $\epsilon_2$ value that saturates 
the electroweak constraint on the $Z$ mass shift.  
To discover a long-lived dark photon model 
in which $\epsilon_2 \simeq 10^{-3}$, however, 
one needs about 3000 fb$^{-1}$ data at the HL-LHC. 
When $\epsilon_2 \simeq {\cal O} (10^{-7})$, 
the LLDP signal approaches 
the value in the conventional LLDP scenario, so that 
it is no longer enhanced by the production channel 
mediated by the $\epsilon_2$ parameter.

\section{LHCb}

\begin{figure}[htbp]
\begin{centering}
\hspace{0.2cm}
\includegraphics[width=0.5\textwidth]{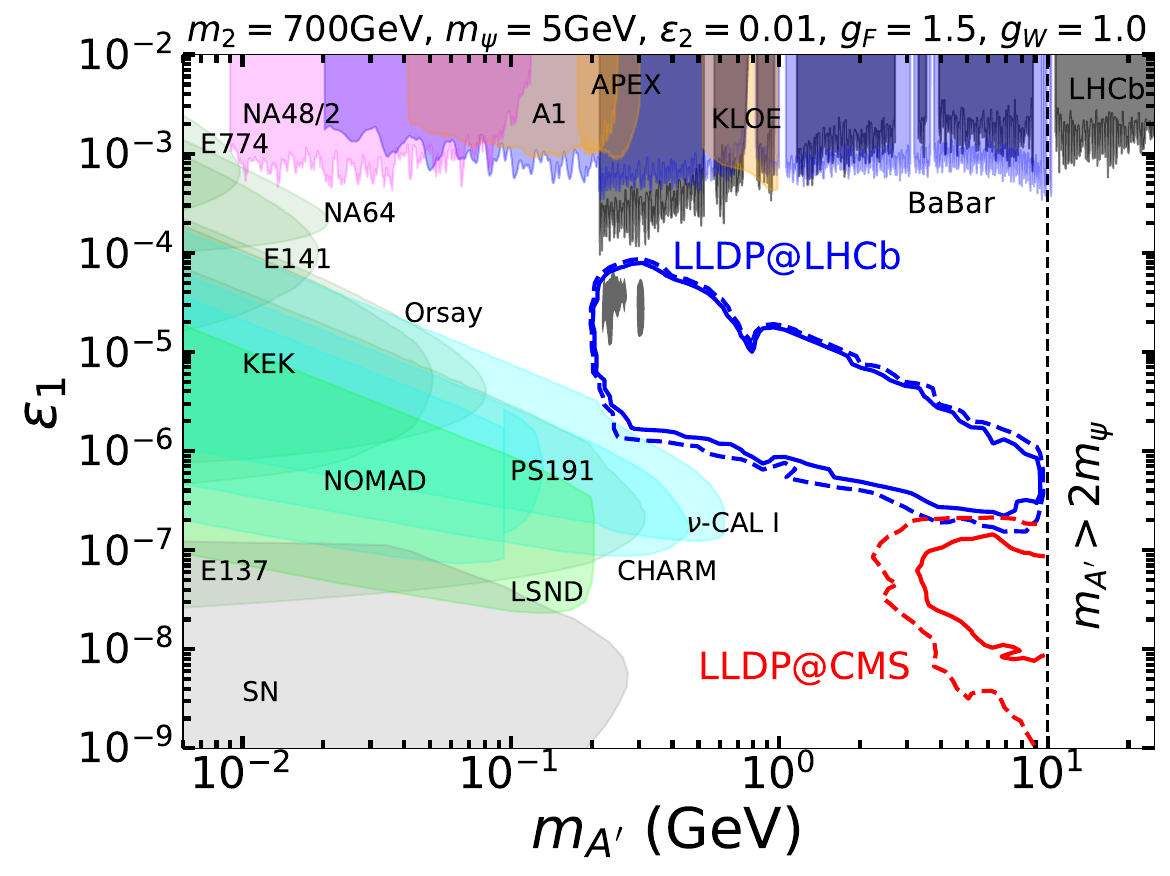}
\caption{LHC current and future sensitivity contours
to the LLDP parameter space spanned by 
$\epsilon_1$ and the dark photon mass. 
We take $m_{\psi} =5$ GeV and $\epsilon_2 = 0.01$.
The blue solid and dashed contours indicate the regions probed by the 
LHCb with $5.5$ $\rm{fb}^{-1}$ and $15$ $\rm{fb}^{-1}$ data respectively. 
The regions probed by the future MTD CMS 
with $250$ ${\rm fb}^{-1}$ and $3000$ ${\rm fb}^{-1}$ data 
are shown as red solid and dashed contours respectively.
The gray islands at $\epsilon_1 \sim (10^{-4}-10^{-5})$ are 
the LHCb exclusion regions for the {conventional} dark photon scenario \cite{Aaij:2019bvg}. 
Various experimental constraints on the {conventional} dark photon scenario 
are shown as color shaded regions. 
}
\label{fig:LHCb}
\end{centering}
\end{figure}

Due to the excellent mass resolution (7-20 MeV)
and vertex resolution ($\sim 10 \,{\rm \mu m}$ on 
the transverse plane) as well as the ability 
for particle identification ($\sim 90\%$ for muons) \cite{Aaij:2015bpa, Benson:2015yzo}, 
the LHCb detector is 
able to discover new elusive particles beyond the SM. %
An upcoming upgrade with an increased luminosity 
and a {more} advanced trigger system with 
only software triggers 
will further improve 
the capability of the LHCb detector to probe new 
physics phenomena, such as LLDPs  \cite{Benson:2015yzo}.
Recently, a search for LLDPs in the kinetic mixing model 
has been carried out at the LHCb 
via displaced muon pairs \cite{Ilten:2016tkc, Aaij:2017rft, Aaij:2019bvg}.
In our model, because LLDPs has a larger production 
cross section at the LHC than the conventional 
LLDP models, the LHCb search \cite{Aaij:2019bvg} 
can probe a much larger parameter space.

To analyze the LHCb constraints, we choose a benchmark point in which 
$m_{\psi} =5$ GeV and $\epsilon_2 = 0.01$, 
while the rest of the parameters take the default values. 
In this benchmark point, the $\psi$ production cross section, 
$\sigma(p p \to \psi \bar{\psi}) \sim 4.3$ pb 
which is dominated by the $Z$-boson exchange channel. 
We use MG5 \cite{Alwall:2014hca} to generate the 
LHC events for each model point on the 
$\epsilon_1$-$m_{A'}$ plane, which are then 
passed to Pythia 8 \cite{Sjostrand:2014zea, Carloni:2010tw, Carloni:2011kk} 
for showering (including showering in hidden sector) 
and hadronization.

We follow the LLDP search criteria in 
Ref. \cite{Aaij:2017rft, Aaij:2019bvg} to analyze the signal.
In particular, we require the transverse distance of 
the dark photon decay vertex of  
$6 \,{\rm mm}<l_T(A')<22\,{\rm mm}$ 
and the pseudo-rapidity of dark photons 
and muons of $ 2 < \eta(A', \mu^{\pm}) < 4.5$. 
These requirements ensure that the displaced vertex is sufficiently separated from the beam line
and registered in the Vertex Locator (VELO) where the dimuon can be reconstructed with good efficiency. 
Furthermore, in order to suppress the background from fake muons, we also require the momentum 
and transverse momentum of muons are greater than $10$ GeV and $0.5$ GeV respectively.

The dominant background includes
the photon-conversion in the VELO, 
muons produced from b-hadron decay chains,  
and pions from $K^0_s$ decays which are misidentified as muons. 
Ref.\ {\cite{Ilten:2016tkc}} estimated the background events
as $B = 25$ for ${\cal L} = 15\, \rm{fb}^{-1}$, 
which is adopted in our analysis and also rescaled for the ${\cal L} = 5.5\, \rm{fb}^{-1}$ case. 
We compute the exclusion region by demanding that 
$S/\sqrt{B} > 2.71$ where $S$ is the signal event number. 

Fig.\ ({\ref{fig:LHCb}}) shows the LHCb exclusion region 
in the parameter space spanned by $\epsilon_1$ and 
the dark photon mass  $m_{A'}$. 
With the current luminosity $5.5$ ${\rm fb}^{-1}$, 
LHCb can probe the parameter space of our model: 
$200\, {\rm MeV} < m_{A'} < 9 \, {\rm GeV}$ 
and $ 2 \times 10^{-7} <  \epsilon_1 < 6 \times 10^{-5}$. 
The exclusion region in the {conventional} dark photon scenario is, 
however, much smaller, 
which is shown as two small gray islands at 
$\epsilon_1 \sim (10^{-4}-10^{-5})$. 
Thus, in our model, 
a significantly larger region of parameter space 
than the {conventional} dark photon model can be probed  
by the current LLDP search at the LHCb. 
A projected limit from the Run 3 data is also computed; 
LHCb can probe the parameter space: 
$200 \, {\rm MeV} < m_{A'} < 10 \, {\rm GeV}$ and $ 10^{-7} <  \epsilon_1 < 10^{-4}$, 
if $15$ $\rm{fb}^{-1}$ integrated luminosity can be 
accumulated in the LHC Run 3 data. 
We note in passing that the shape of the exclusion contours is primarily due to the 
detector cut on the dark photon decay length: a smaller $\epsilon_1$ value 
is needed in the larger dark photon mass region so that the dark 
photon has the desired decay width to disintegrate in the VELO region. 
Also the dip at $m_{A'} \simeq 0.8$ GeV is due to 
the $\omega$ resonance which suppresses the BR($A' \to \mu^+\mu^-$). 
Fig.\ ({\ref{fig:LHCb}}) also shows the exclusion limits on 
the conventional
dark photon from various experiments; 
the limits are taken from the Darkcast package \cite{darkcast}.

Fig.\ ({\ref{fig:LHCb}}) also shows the sensitivities from 
the future MTD CMS detector via the time delay measurement. 
As mentioned before, the time delay signal from the final state leptons becomes more significant 
if the LLDPs have long lifetime and move non-relativistically. 
Therefore, the timing detector probes the heavy dark photon mass 
region with a smaller mixing parameter $\epsilon_1$ 
which are currently almost inaccessible at the LHCb. 
In particular, with the luminosity of $250 \, \rm{fb}^{-1}$ at the HL-LHC, 
the MTD CMS detector can probe the parameter space: 
$m_{A'} > 3.3 $ GeV 
and $ 10^{-8} < \epsilon_1 < 10^{-7}$. 
An even larger parameter space in our model: 
$m_{A'} > 2.0 $ GeV and $ 10^{-9} < \epsilon_1 < 2\times 10^{-7}$, 
can be reached with $3000 \, \rm{fb}^{-1}$ data 
accumulated at the HL-LHC. 
Interestingly, this MTD CMS sensitivity 
region partly overlaps with 
the LHCb sensitivity region with 15 fb$^{-1}$ data.
Thus, 
if the LLDP is discovered in this overlapped region, 
the timing detector can be used to verify the LHCb results.  
We note that, in the region of $m_{A'} > 2 m_{\psi}$, 
the dark photon will dominantly decay into ${\psi}$ 
so that it
can no longer be searched for in the 
visible channel by the LHCb detector and the 
future precision timing detectors.

\section{Summary}

We construct a long-lived dark photon model which 
has an enhanced dark photon collider signal. 
We extend the standard model by a hidden sector 
which has two gauge bosons and one Dirac fermion $\psi$; 
the two gauge bosons interact with the SM sector via 
different Stueckelberg mass terms. 
The GeV-scale dark photon $A'$ interacts with the SM fermions via 
a very small Stueckelberg mass term (parametrized by 
the dimensionless quantity $\epsilon_1$) such that it 
has a macroscopic decay length which can lead to a 
displaced vertex or a time delay signal at the LHC. 
The TeV-scale $Z'$ boson 
interacts with the SM via a relatively 
larger mass term (parametrized by 
the dimensionless quantity $\epsilon_2$). 
Because the dark photon $A'$ is mainly produced 
at the LHC via the $\psi$ dark radiation 
processes in which the effective coupling strength 
is of the size of $\epsilon_2$ 
the LHC signal of $A'$ 
is thus enhanced significantly.

Various experimental constraints on the model 
are analyzed, 
including the electroweak constraint on the $Z$ boson mass shift, 
the constraint from the $Z$ invisible decay, 
LHC constraints, 
collider constraints on millicharge, 
and cosmological constraints on millicharge.
The electroweak constraint on the $Z$ mass turns out to be 
the most stringent one, 
which leads to an upper bound $\epsilon_2 \lesssim 0.036$, 
in the parameter space of interest.

Two types of LHC signals from the LLDP in our model are investigated: 
the time delay signal measured by the precision timing detectors 
at the HL-LHC, 
and the current LHCb searches on LLDPs. 
If the LLDP is produced non-relativistically at the LHC, 
it has a significant time delay $\Delta t$, 
which can be measured by the precision timing detectors. 
Under the detector cut $\Delta t > 1.2$ ns, 
the SV and PU backgrounds are found to be negligible. 
The parameter space 
of 3 GeV$< m_{A'}, m_{\psi} <$ 30 GeV 
in our model is found to be probed by the timing 
detector with 250 ${\rm fb^{-1}}$ data 
at the HL-LHC.

Due to the different search strategy, the current LHCb 
analysis is more sensitive to the lighter dark photon mass 
than the time delay searches. 
We found that the parameter space probed by the current LHCb 
analysis is much larger in our model than the conventional 
dark photon model investigated in the LHCb experimental analysis. 
A comparison between the LHCb search and the 
time delay search is also made; 
they typically probe 
different regions of the parameter space 
but can overlap in some small regions.

We note that a similar model as ours 
can be constructed by introducing two kinetic mixing 
parameters, of magnitude ${\cal O}(10^{-2})$ 
and ${\cal O}(10^{-7})$, which are responsible 
for dark photon production and decay processes, 
respectively.

\section{Acknowledgement}

We thank Jinhan Liang, Lei Zhang, and T.\ C.\ Yuan 
for helpful discussions 
and correspondence. 
The work is supported in part  
by the National Natural Science Foundation of China under Grant Nos.\ 
11775109 and U1738134.

\appendix

\section{Decay and Lifetime of Dark Photon $A'$} 
\label{app:decay}

\begin{figure*}[htbp]
\begin{centering}
\includegraphics[width=0.45\textwidth]{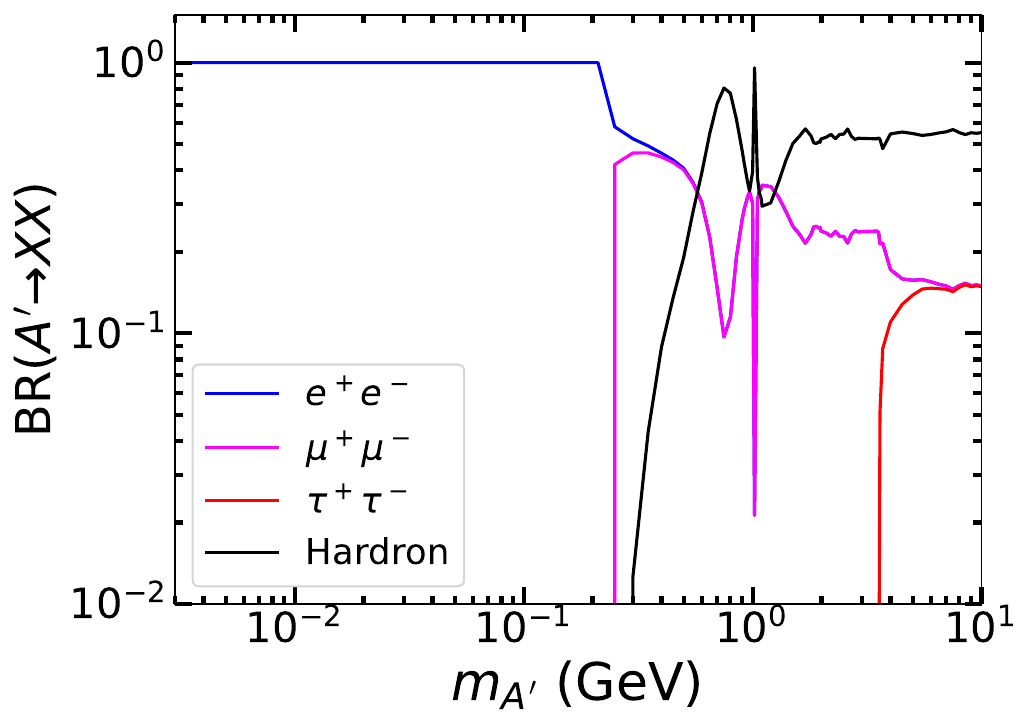}
\includegraphics[width=0.45\textwidth]{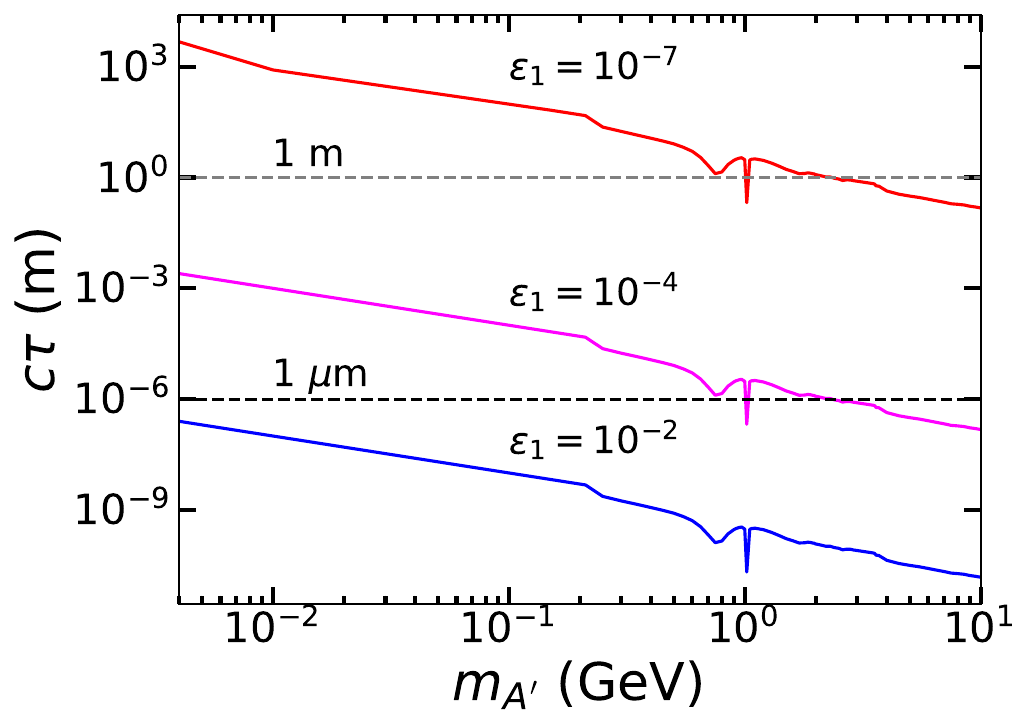} 
\caption{Left panel: 
Branching ratio of the dark photon.
Right panel: 
Proper lifetime of the dark photon with various values of $\epsilon_1$. 
The dashed black line on the right panel figure 
indicates the criterion for the 
LHC prompt decay: $c\tau \leq 1$ $\mu m$. 
}
\label{fig:BR-LT}
\end{centering}
\end{figure*}

The dark photon leptonic decay width is given by 
\be
\label{eq:Z1toll}
\Gamma(A'\to l^+l^-) =
\frac{m_{A'} }{12 \pi } \sqrt{1-4 \frac{m_l^2}{m_{A'}^2}} \left[ \left(1-4 \frac{m_l^2}{m_{A'}^2}\right)  |a^l_2|^2 + \left(1+ 2 \frac{ m_l^2}{m_{A'}^2} \right) |v^l_2|^2 \right] \,,
\ee
where $v^l_2$ and $a^l_2$ are the vector and axial-vector couplings between the dark photon and the leptons 
which are given in Eqs.~(\ref{eq:vcouplings},\ref{eq:acouplings}). 
The hadronic decay width can be computed by
\be
\label{eq:Z1tohad}
\Gamma (A' \to {\rm hadrons}) = \Gamma( A' \to \mu^+\mu^-) \,R(m_{A'}^2), 
\ee
where $R(m^2_{A'}) \equiv \sigma(e^+e^-\to {\rm hadrons})/\sigma(e^+e^-\to \mu^+ \mu^-)$ 
takes into account the effects of the dark photon mixing with the QCD vector mesons 
and is taken from Ref.\ \cite{Tanabashi:2018oca}. 
The $\Gamma(A'\to \psi\bar\psi)$ can be computed by replacing the 
couplings and mass for leptons in Eq.\ (\ref{eq:Z1toll}) 
with the ones for $\psi$. We note that below a few hundred MeV, 
dark photon decays into the $e^+e^-$ and $\mu^+\mu^-$ pairs. 
In our analysis, since we consider $m_{A'} < 2 m_\psi$, 
the $\Gamma(A'\to \psi\bar\psi)$ is kinematically forbidden. 

Fig.~(\ref{fig:BR-LT}) shows the SM branching ratios of the dark photon 
decay 
and its proper lifetime with different $\epsilon_1$ values. 
We note that a particle is considered as a long-lived particle 
in particle colliders if its decay length is larger 
than the detector spatial resolution
which can vary from ${\cal O}(10)\mu$m to ${\cal O}(10)$mm 
depending on the detectors.

\section{Approximated couplings}
\label{app:couplings}
 
Here we provide approximated expressions of the  
vector and axial-vector couplings for the three massive 
gauge bosons, 
where $m_1 = 1$ GeV, $m_2 = 700$ GeV, 
$\epsilon_1\ll 1$ and $\epsilon_2 \ll 1$. 
The approximated expressions of the 
vector and axial-vector couplings for the $Z'$ boson are 
\bea
\label{eq:vcouplingsZp}
v^f_1 &\simeq& \left( -0.18\, T^{3 }_f + 0.35\, Q_f \right) \epsilon_2, \\ 
\label{eq:acouplingsZp}
a^f_1 &\simeq&   -0.18\, T^{3 }_f \, \epsilon_2 \,, \\
\label{eq:couplingsZp} 
v^\psi_1 &\simeq& g_{W}  +2.0\times 10^{-6}\, \epsilon_1 \epsilon_2  \, g_{F}. 
\eea
The approximated expressions of the $A'$ boson couplings are 
\bea
\label{eq:vcouplingsAp}
v^f_2 &\simeq& \left( 2.1 \times 10^{-5} \, T^{3 }_f + 0.27 \,Q_f \right)\epsilon_1 , \\ 
\label{eq:acouplingsAp}
a^f_2 &\simeq& 2.1 \times 10^{-5} \, T^{3 }_f \, \epsilon_1, \\
\label{eq:couplingsAp} 
v^\psi_2 &\simeq&  g_{F} -0.78 \, \epsilon_1 \epsilon_2\, g_{W}  . 
\eea
The $A'$ boson has a much larger vector coupling to the 
SM fermions than the axial-vector coupling, and the 
vector coupling is proportional to the electric charge 
when neglecting the small contribution from the isospin.  
The $A'$ boson is called ``dark photon'' due to the fact 
it has a photon-like interaction to 
the SM fermions.

The approximated expressions of the $Z$ boson couplings are 
\bea
\label{eq:vcouplingsZ}
v^f_3 &\simeq& v^{f0}_{Z} 
+ \left(  0.041 \, \epsilon_1^2 - 1.4 \times 10^{-3} \, \epsilon_2^2 \right) T^{3 }_f
 -( 0.18 \,\epsilon_1^2  + 0.034 \,\epsilon_2^2 \ )  Q_f ,\\ 
\label{eq:acouplingsZ}
a^f_3 &\simeq& a^{f0}_{Z} + \left(  0.041 \, \epsilon_1^2 - 1.4 \times 10^{-3} \, \epsilon_2^2 \right) T^{3 }_f \\
\label{eq:couplingsZ} 
v^\psi_3 &\simeq&  0.48 \,  \epsilon_2\, g_{W}  -  5.8 \times 10^{-5} \,\epsilon_1 \, g_{F} ,
\eea
where $v^{f0}_{Z}$ and $a^{f0}_{Z}$ are the SM values,  
\bea
v^{f0}_{Z} &=& {e \over 2 s_W c_W} (T^{3 }_f -  2 s^2_W Q_f ) , \\
a^{f0}_{Z} &=& {e \over 2 s_W c_W} T^{3 }_f ,
\eea
where the weak rotation angle is given by 
$\tan\theta_W = g'_{\rm SM}/g$ 
and $e=g s_W = g'_{\rm SM} c_W$.

\end{document}